\documentstyle[12pt,epsfig,rotating]{article}

\newcommand{\be}{\begin{equation}}
\newcommand{\ben}{\begin{subequations}}
\newcommand{\een}{\end{subequations}}
\newcommand{\beq}{\begin{eqalignno}}
\newcommand{\eeq}{\end{eqalignno}}
\newcommand{\ee}{\end{equation}}

\newcommand{\mchi}{\mbox{$m_{\tilde {\chi}_1^0}$}}
\newcommand{\lsp}{\mbox{$\tilde {\chi}_1^0$}}
\newcommand{\Ochi}{\mbox{$\Omega_{\tilde \chi} h^2$}}
\newcommand{\tanb}{\mbox{$\tan \! \beta$}}
\newcommand{\sto}{\mbox{$\tilde{t}_1$}}
\newcommand{\stos}{\mbox{$\tilde{t}_1^*$}}

\newcommand{\msto}{\mbox{$m_{\tilde{t}_1}$}}
\def\lsim{\:\raisebox{-0.5ex}{$\stackrel{\textstyle<}{\sim}$}\:}
\def\gsim{\:\raisebox{-0.5ex}{$\stackrel{\textstyle>}{\sim}$}\:}

\begin{document}  

\begin{flushright}
PM/99--53\\
TUM-HEP-363-99
\end{flushright}

\vspace{1cm}

\begin{center}

{\large\sc {\bf Light Scalar Top Quarks and Supersymmetric Dark Matter}}

\vspace{1cm}

{\sc C\'eline Boehm$^1$, Abdelhak Djouadi$^1$ and Manuel Drees$^2$}

\vspace{0.5cm}

$^1$ Laboratoire de Physique Math\'ematique et Th\'eorique, UMR5825--CNRS,\\
Universit\'e de Montpellier II, F--34095 Montpellier Cedex 5, France.

\vspace*{2mm}

$^2$ Physik Department, Technische Universit\"at M\"unchen, \\
James Franck Str., D--85748 Garching, Germany
\end{center}

\vspace{2cm}

\begin{abstract}

\noindent 
A stable neutralino \lsp, assumed to be the lightest supersymmetric particle,  
is a favored particle physics candidate for the cosmological Dark Matter. We 
study co--annihilation of the lightest neutralino with the lighter scalar top 
quark \sto. We show that for natural values of the neutralino mass, $\lsim 300$
GeV, the \lsp--\sto\ mass difference has to exceed $\sim 10$ to 
30 GeV if \lsp\ is to contribute significantly to the Dark Matter. Scenarios 
with smaller mass splitting, where \sto\ is quite difficult to detect at 
collider experiments, are thus cosmologically disfavored. On the other hand, 
for small \sto--\lsp\ mass splitting, we show that co--annihilation allows 
very large neutralino masses, $\mchi \sim 5$ TeV, without ``overclosing'' the 
Universe.
\end{abstract}

\newpage

There is convincing evidence \cite{1} that most matter in the Universe
is dark (non--luminous),
\be \label{e1}
0.2 \lsim \Omega_{\rm DM} \leq 1
\ee
where $\Omega_{\rm DM}$ is the Dark Matter (DM) density in units of the
critical density, so that $\Omega = 1$ corresponds to a flat
Universe. On the other hand, analyses of Big Bang nucleosynthesis
\cite{2} imply that most DM is non--baryonic (although dark baryons
probably exist as well). 

One of the favorite particle physics candidates for DM is the lightest
neutralino \lsp\ \cite{3}, assumed to be the lightest supersymmetric
particle (LSP). It is stable if R-parity is conserved \cite{fayet}; this 
is also a sufficient (although not necessary) condition for avoiding very 
fast nucleon decay in supersymmetric theories. The LSP makes an attractive
DM candidate since the primary motivation for its introduction comes
from particle physics arguments \cite{4}: supersymmetry stabilizes the 
huge hierarchy between the weak and Grand Unification scales against 
radiative corrections, and if it is broken at a sufficiently high scale, 
it allows to understand the origin of the hierarchy in terms of radiative 
breaking of the Standard Model (SM) electroweak ${\rm SU_L(2) \times U(1)_Y}$ 
gauge symmetry; furthermore it allows for a consistent unification of the 
gauge couplings.

Supersymmetric contributions to DM then come as extra bonus, and for wide 
regions of parameter space, the LSP relic density falls in the preferred 
range eq.(\ref{e1}). This is true in particular if the LSP is mostly a
superpartner of the ${\rm U(1)_Y}$ gauge boson, i.e. bino--like, and if 
both \mchi\ and the masses of ${\rm SU(2)}$ singlet scalar leptons fall 
in the natural range below a few hundred GeV \cite{5} (but above \cite{6} 
the mass range excluded by the LEP experiments).

The previous statement assumes that \lsp\lsp\ annihilation reactions are 
the only processes that change the number of superparticles at
temperatures around $T_F \simeq \mchi/20$, where the neutralino \lsp\ 
decouples from the plasma of SM particles. It has been 
known for some time \cite{9} that this is not true if the mass splitting 
between the LSP and the next--to--lightest supersymmetric particle 
$\tilde{P}$ is small. In this case, reactions of the type
\be \label{e2}
\lsp + X \leftrightarrow \tilde{P} + Y,
\ee
where $X,Y$ are SM particles, occur much more frequently at a temperature
$T \sim T_F$ than \lsp\lsp\ annihilation reactions do. The rate of the 
latter kind of process is proportional to two powers of the Boltzmann 
factor $ \exp(-\mchi/T_F) \simeq \exp(-20)$, whereas for $\mchi \simeq
m_{\tilde P}$ the rate for reaction (\ref{e2}) is linear in this
factor. These reactions will therefore maintain {\em relative} equilibrium 
between the states \lsp\ and $\tilde{P}$ until long after all superparticles 
decouple from the Standard Model plasma.

The total number of superparticles can then not only be changed by
\lsp\lsp\ annihilation, but also by the ``co--annihilation'' processes 
\be \label{e2p}
\lsp + \tilde{P} \leftrightarrow X + Y \; {\rm and} \;  \tilde{P} + 
\tilde{P}^{(*)} \leftrightarrow X + Y. 
\ee
Eventually all particles $\tilde{P}$ and $\tilde{P}^*$ will decay into \lsp\
(plus SM particles). In order to compute today's LSP relic density, we
therefore only have to solve the Boltzmann equation for the sum
$n_{\rm SUSY}$ of densities $n_i$ of all relevant species of
superparticles. In this sum contributions from reactions (\ref{e2})
cancel, since they do not change the total number of superparticles.
One thus has \cite{9}
\begin{eqnarray} \label{boltz}
\frac {d n_{\rm SUSY}} {d t} &&= \, - 3 H n_{\rm SUSY} - \sum_{i,j} \langle
\sigma_{ij} v \rangle \left( n_i n_j - n_i^{\rm eq} n_j^{\rm eq} \right)
\nonumber \\
&&= \, - 3 H n_{\rm SUSY} - \langle \sigma_{\rm eff} v \rangle \left(
n^2_{\rm SUSY} - n^{{\rm eq}^2}_{\rm SUSY} \right).
\end{eqnarray}
Here, $H$ is the Hubble parameter, $\langle \dots \rangle$ denotes
thermal averaging, $v$ is the relative velocity between the two
annihilating superparticles in their center--of--mass frame, and the
superscript ``eq'' indicates the equilibrium density. In the second
step we made use of the fact that, as argued above, all relevant
heavier superparticles maintain relative equilibrium to the neutralino
LSP until long after the temperature $T_F$. This allowed us to sum all 
superparticle annihilation processes into an ``effective'' cross section;
schematically \cite{9}
\begin{eqnarray} 
\label{e3}
\sigma_{\rm eff} \propto  \; 
g_{\tilde{\chi}
\tilde{\chi}} \sigma(\lsp\lsp) + g_{\tilde{\chi} \tilde{P}} B_{\tilde P}
\sigma(\lsp \tilde{P}) 
+ g_{\tilde{P}\tilde{P}} \left( B_{\tilde P} \right)^2\sigma(\tilde{P} 
\tilde{P}^{(*)}) .
\end{eqnarray}
where the $g_{ij}$ are multiplicity factors, and
\be \label{e4}
B_{\tilde P} = ( m_{\tilde P}/ m_{\tilde \chi_1^0})^{3/2} e^{-(m_{\tilde P} 
- m_{\tilde \chi_1^0})/T }
\ee
is the temperature dependent relative Boltzmann factor between the
$\tilde{P}$ and \lsp\ densities. The final LSP relic density \Ochi,
where $h = 0.65 \pm 0.15$ is the scaled Hubble constant, is then
essentially inversely proportional to $\langle \sigma_{\rm eff} v
\rangle$ at $T_F \simeq \mchi/20$. Co--annihilation can therefore
reduce the LSP relic density by a large factor, if $\delta m \equiv
m_{\tilde P} - \mchi \ll \mchi$ and $\sigma(\lsp\tilde{P}) +
\sigma(\tilde{P} \tilde{P}^{(*)}) \gg \sigma(\lsp\lsp)$. This is true
in particular if \lsp\ is a light, $\mchi < M_W$, higgsino \cite{5,7}
or SU(2)--gaugino \cite{8}. More recently it has been pointed out
\cite{10,11} that co--annihilation with light sleptons can reduce the
relic density of a bino--like LSP by about one order of magnitude.

In this letter we study co--annihilation of neutralinos with the lighter 
scalar top (stop) eigenstate \sto. Compared to the other squarks, \msto\ is
reduced \cite{4} by contributions of the large top quark Yukawa coupling to
the relevant renormalisation group equations, as well as by mixing
between SU(2) doublet and singlet stops. While we do not know of any
model that predicts $\msto \simeq \mchi$, a close mass degeneracy is
possible in many models, e.g. in the popular minimal Supergravity (mSUGRA)
model \cite{4}. Moreover, scenarios with small \sto--\lsp\ mass splitting
are of great concern for experimenters, since \sto\ decays then
release little visible energy, making \sto\ production very difficult
to detect at both $e^+e^-$ \cite{13} and hadron \cite{14}
colliders.

In contrast to the cases mentioned earlier, for $\tilde{P}=\sto$ it is
not entirely obvious that reactions of the type (\ref{e2}) will indeed
be much faster than \lsp\lsp\ annihilation processes. In the absence
of flavor mixing, one would have to chose $X=W, Y=b$ or vice versa. However, 
for a temperature $T < M_W$, the $W$--density is itself quite small, so
reaction (\ref{e2}) would be much faster than \lsp\lsp\ annihilation
only for \mchi\ significantly above $M_W$. On the other hand, most
supersymmetric models predict some amount of flavor mixing in the squark 
sector, even if it is absent at some high energy scale. As a result, for 
small $\delta m$ the dominant \sto\ decay mode is usually its flavor changing
2--body decay into $\lsp+c$ \cite{15}. For \sto\ masses of current
experimental interest the dominant contribution to (\ref{e2})
therefore comes from $X=c, Y=$ nothing, i.e. (inverse) \sto\ decay. If
the effective $c \sto \lsp$ coupling is suppressed by a small mixing
angle $\epsilon$, the condition that (\ref{e2}) is much faster than
\lsp\lsp\ annihilation reads
\be \label{e5}
\epsilon^2 e^{-\delta m/T_F} \gg \alpha e^{-m_{\tilde \chi_1^0}/T_F},
\ee
where the extra factor of $\alpha \sim 0.01$ occurs since we are
comparing $2 \leftrightarrow 1$ reactions with $2 \leftrightarrow 2$
processes. For $\delta m \lsim T_F \sim \mchi/20$ we then only
need $\epsilon > e^{-10} \simeq 5 \cdot 10^{-5}$. In what follows we
will assume that this is true, or that \lsp\ is sufficiently heavy
that $\lsp+W^+ \leftrightarrow \sto+ \bar{b}$ is fast. 

Another property of the stop is that it has strong interactions. A
leading order calculation of $\sigma(\lsp\sto)$ and
$\sigma(\sto\sto^{(*)})$ will therefore not be very
reliable. Unfortunately a full higher order calculation is highly
nontrivial, since one would need to include finite temperature effects
(e.g. in order to cancel Coulomb singularities in the
non--relativistic limit). We expect these unknown higher order QCD
corrections to be more important than the contributions of higher
partial waves. In the calculation of the cross sections $\sigma(\lsp\sto)$ 
and $\sigma(\sto \tilde{t}_1^{(*)})$ we therefore only include the leading,
$S-$wave contribution; however, the $P-$wave contributions to
\lsp\lsp\ annihilation process \cite{5} are included. Our co--annihilation
cross sections will thus only be accurate to a factor of 2 or
so. Due to the exponential dependence of $\sigma_{\rm eff}$
on $\delta m$, see eqs.(\ref{e3}--\ref{e4}), the bounds on the
\sto--\lsp\ mass splitting that will be inferred from upper or lower
bounds on \Ochi\ should nevertheless be fairly accurate.

The existence of unknown, but probably large, higher order corrections
also means that we can ignore all \sto\ annihilation reactions that
involve more than the minimal required number of electroweak gauge
couplings. However, we treat the top and bottom quarks Yukawa couplings 
on the same footing as the strong coupling (the latter Yukawa coupling
will be large only for $\tanb \sim m_t/m_b$). Altogether we therefore 
computed the cross sections for the following processes:
\begin{eqnarray} \label{e6} 
\lsp \sto & \rightarrow  & t~g, \ t~H_i^0, \ b~H^+ ; 
\label{e6a} \nonumber \\
\sto \sto & \rightarrow & t~t ;
\label{e6b} \nonumber \\
\sto \stos & \rightarrow & g~g, \ H_i^0 ~ H_j^0, \ H^+ ~ H^-, \ b ~
\bar{b}, \ t ~ \bar{t} \ ,  
\label{e6c}
\end{eqnarray} 
where $H_i^0\equiv h,H,A$ is one of the three neutral Higgs bosons of the 
minimal supersymmetric Standard Model (MSSM) \cite{17}. The cross sections for
$\lsp \stos$ and $\stos \stos$ annihilation are identical to those in
the first and second lines of eq.(\ref{e6}), respectively. We have performed 
two independent calculations of these cross sections. One calculation was
based on trace techniques and the usual polarization sum for external
gluons; here the non--relativistic limit (to extract the $S-$wave
contribution \cite{3}) was only taken at the end. The second method
uses helicity amplitudes \cite{5}; in this case the non--relativistic
limit can already be taken at the beginning of the calculation. [Note
that the cross sections for $\sto \sto^* \rightarrow H_i^0 ~ g$ vanish
in this limit.] Explicit expressions for these cross sections will be
published elsewhere. 
[Note that in their analytical expressions, the authors of Refs.\cite{10,11} 
do not keep the mass of the relevant SM fermion $f$; these terms are 
irrelevant for their case $f=\tau$, whereas we have to keep a finite value 
for the top quark mass, $m_t \neq 0$. Ref.\cite{10} also did not include
$\tilde{f}_L-\tilde{f}_R$ mixing, which in our case is crucial for
obtaining a light \sto. In the relevant limit we agree with Ref.\cite{10}. 
We disagree with Ref.\cite{11} for the $tg$ (or $\tau \gamma$) final
state, by a factor $\mchi/\sqrt{s}$.]

Our calculation of the relic density closely follows ref.\cite{9}. In
particular, we keep the temperature dependence (\ref{e3}) of
$\sigma_{\rm eff}$ when computing the ``annihilation integral''
(essentially the integral of the Boltzmann equation for $T>T_F$).

We use a variant of the minimal Supergravity model \cite{4} for our
numerical analysis. In particular, we assume a common gaugino mass,
a common sfermion mass $m_0$, and a common trilinear soft breaking
parameter $A_0$ at the Grand Unification scale $M_X = 2 \cdot 10^{16}$
GeV. However, we allow the soft breaking masses of the two Higgs
doublets to differ from $m_0$. In practice, this means that we keep the
higgsino mass parameter $\mu$ and the mass $m_A$ of the CP--odd Higgs
boson as free parameters at the weak scale. The final free parameter
is the ratio \tanb\ of vacuum expectation values of the two Higgs
fields.

For illustration, we take $\mu = - 2 M_2$, where $M_2 \simeq 2 \mchi$ is the 
SU(2) gaugino mass. This implies that the LSP is bino--like, which is the
most natural choice for this type of model \cite{18}. It is also
conservative, since a higgsino--like LSP will have larger couplings to
the (s)top, and hence even larger co--annihilation cross sections
eq.(\ref{e6a}). We also chose a large sfermion mass, $m_0 = 2 M_2$. In
the absence of co--annihilation this choice is usually incompatible
\cite{5} with the upper bound on the LSP relic density, which we
conservatively take as $\Ochi \leq 0.5$. 

\begin{figure}[htb]
\vspace*{-0.6mm}
\hspace*{15mm}
\begin{turn}{-90}%
\mbox{
\epsfig{file=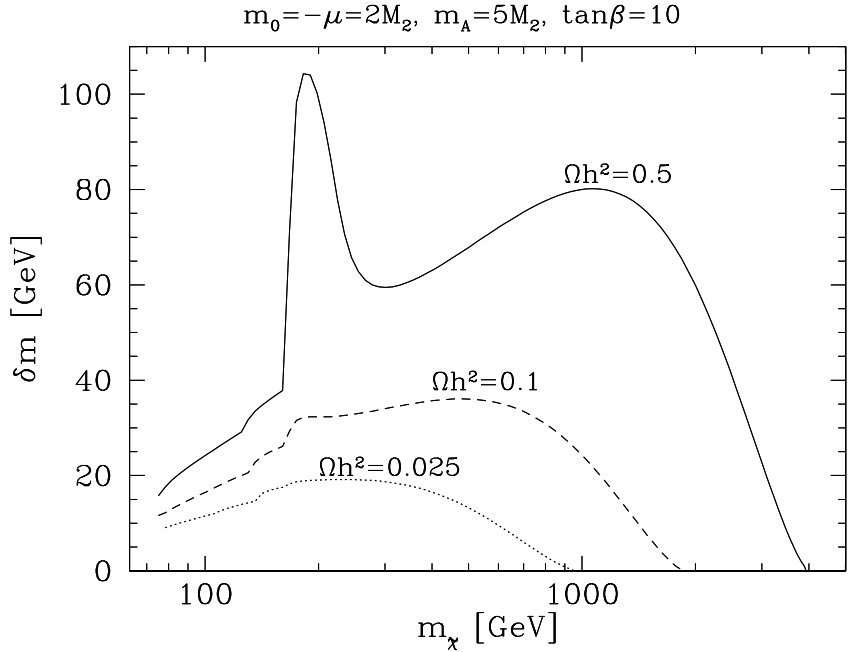,width=0.55\textwidth} }
\end{turn}
\vspace*{-1.mm}
\caption{Contours of constant $\Ochi=0.5$ (solid), $0.1$ (dashed) and
$0.025$ (dotted) in the $(\mchi,\delta m)$ plane, where $\delta m =
\msto - \mchi$. We took $\mu, \ m_0$ and $m_A$ to be fixed multiples
of $M_2 \simeq 2 \mchi$, as indicated, whereas $\tanb=10$ has been
kept fixed. The parameter $A_0$ varies between about $2.5 m_0$ and
$3.2 m_0$, with larger $A_0$ values corresponding to smaller values of
$\delta m$. }
\end{figure}

In Fig.~1 we show contours of constant \Ochi\ in the $(\mchi, \delta
m)$ plane for a scenario with moderate \tanb\ and a very heavy Higgs
spectrum, $m_A = 5 M_2$. This latter choice implies that the only
Higgs boson relevant for the calculation of the LSP relic density is
the light CP--even scalar $h$, with mass $m_h \leq 130$ GeV. This is a
conservative scenario in the sense that it minimizes the number of
final states contributing in eqs.(\ref{e6}), and also leads to a small
\lsp\lsp\ annihilation cross section. We see that scenarios with very
large $\delta m$ are indeed excluded by the upper bound on \Ochi. The
peak in the contour $\Ochi=0.5$ at $\mchi \simeq m_t$ is due to
$\lsp\lsp \rightarrow t \bar t$, which has a sizable $S-$wave cross
section if \sto\ is not too heavy and \mchi\ is not much above
$m_t$. The much smaller bumps at $\mchi \simeq 130$ GeV are due to
$hh$ final states becoming accessible.

On the other hand, for very small values of $\delta m$ and \mchi\ in
the range indicated by naturalness arguments ($\lsim 0.3$ TeV,
corresponding to a gluino mass $m_{\tilde g} \lsim 2$ TeV), we find
that the LSP cannot contribute significantly to the solution of the
Dark Matter puzzle, since its relic density is too small. In
particular, one needs \cite{3} $\Ochi > 0.025$ for \lsp\ to form
galactic haloes. We see that even for the present very conservative
choice of parameters one needs a \sto--\lsp\ mass splitting of at
least 9 to 19 GeV (6 to 10\%) to satisfy this lower bound on
\Ochi. This mass splitting is large enough for standard \sto\ search
methods at $e^+e^-$ colliders \cite{13,20} to have reasonably high
efficiency. If we require that \Ochi\ lies in the currently favored
``best fit'' range between about 0.1 and 0.2, $\delta m$ has to be
between 11 and 33 GeV. Unfortunately this is still not high enough for
current \sto\ search strategies at the Tevatron \cite{14} to be
sensitive. 

So far we have focused on LSP masses in the range favored by
naturalness arguments. It is sometimes claimed \cite{19} that the
upper bound on \Ochi\ implies that LHC experiments must find
superparticles if the MSSM is correct and \lsp\ is
bino--like. Unfortunately this is not true; for $\delta m \rightarrow
0$ an LSP mass up to 4 TeV, corresponding to a gluino mass in excess
of 20 TeV, cannot be excluded from this cosmological
argument. [As noted above, our estimates for \sto\
annihilation cross sections are not very reliable. However, even if we
over--estimated them by a factor of 2, the bound on \mchi\ would only
be reduced by a factor of $\sqrt{2}$, and would thus still allow
sparticle masses far above the range to be covered by the LHC.]

In Fig.~2 we show analogous results for a light spectrum of Higgs
bosons and large \tanb, where the bottom Yukawa coupling is sizable;
the choice $m_A = 0.35 M_2 \simeq 0.7 \mchi$ ensures that all Higgs
pair final states will be accessible for $\mchi > 100$ GeV. However,
we keep the previous (large) values for $|\mu|$ and
$m_0$. Nevertheless we see that for natural values of \mchi, requiring
$\Ochi > 0.025$ now implies $\delta m > 20$ GeV. Moreover, the LSP
makes a good DM candidate, i.e. $\Ochi \sim 0.1$, only for $\delta m
\gsim 40$ GeV; this is sufficiently large to permit \sto\ searches at
the Tevatron \cite{13,21}. Finally, for $\delta m \rightarrow 0$,
cosmology now allows an LSP mass up to 6 TeV, corresponding to a
gluino mass of about 30 TeV! Obviously the upper bound on $\delta m$
that follows from $\Ochi>0.025$ for natural values of \mchi, as well
as the absolute upper bound on \mchi\ that follows from $\Ochi \leq
0.5$, are even higher if we chose smaller values for $m_0$ and/or
$|\mu|$.

\begin{figure}[htbp]
\vspace*{-.6mm}
\hspace*{15mm}
\begin{turn}{-90}%
\mbox{
\epsfig{file=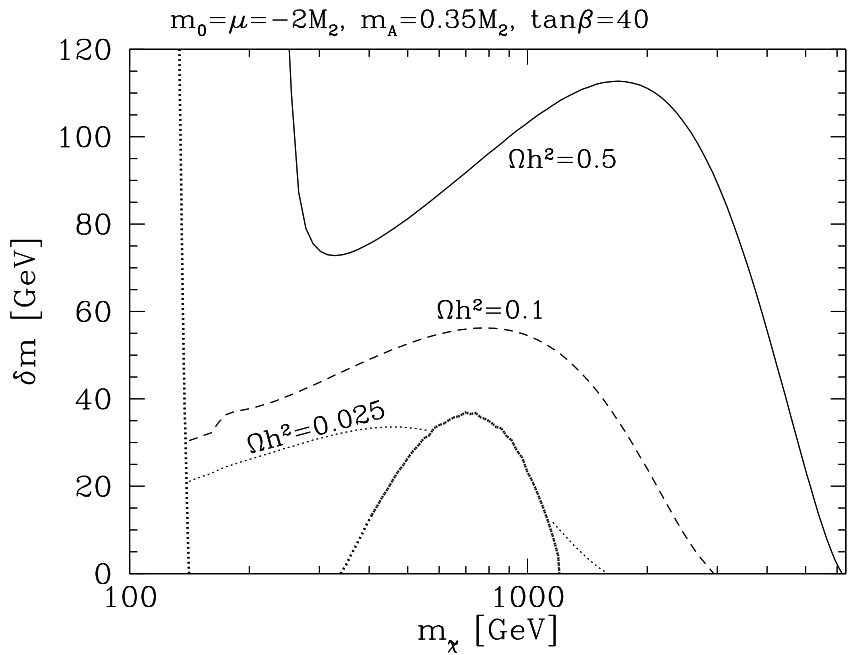,width=0.55\textwidth} }
\end{turn}
\vspace*{-1.mm}
\caption{As in Fig.~1, except that we took a large value of \tanb\ and
a light Higgs boson spectrum. The regions below and to the left of the
heavy dotted lines are excluded by Higgs boson searches at LEP.}
\end{figure}

In conclusion, we have shown that scenarios with very small
\sto--\lsp\ mass splitting would permit an LSP mass of several TeV
without ``overclosing'' the Universe. This shows once again \cite{5}
that the upper bound on the LSP relic density does not guarantee that
LHC experiments will detect superparticles, even if the MSSM is
correct; of course, (third generation) superparticles with masses out
of the reach of the LHC can hardly be argued to be ``natural''. On the
other hand, for \lsp\ and \sto\ masses of present experimental
interest, and indeed for the entire natural range of these masses,
\lsp\ cannot contribute significantly to the Dark Matter in the
Universe unless the \sto--\lsp\ mass difference is large enough for
conventional \sto\ search strategies at $e^+e^-$ colliders to be
effective. This does not imply that collider searches for \sto\ nearly
degenerate with \lsp\ \cite{22} should not be continued; a positive
signal would definitely exclude \lsp\ as DM candidate, which is not
easy to accomplish with cosmological Dark Matter searches. However,
since Dark Matter is known to exist, for natural values of \mchi\ a
very small \sto--\lsp\ mass splitting would require physics beyond the
MSSM.

\bigskip

\noindent {\bf Acknowledgments:} \\
We thank T. Falk and K. Pallis for clarifications regarding Refs.~\cite{10} 
and \cite{11}, respectively.
MD thanks the CNRS for support and the members of the LPMT Montpellier
for their hospitality. This work is partially supported by the
French GDR--Supersym\'etrie.

\end{document}